# The impact of Hubbard- and van der Waals-corrections on the DFT calculation of epsilon-zeta transition pressure in solid oxygen

**Le The Anh[1*], Masahiro Wada[2], Hiroshi Fukui[2], and Toshiaki Iitaka[1]**

[1]Computational Astrophysics Laboratory, RIKEN, 2-1 Hirosawa, Wako, Saitama 351-0198, Japan
[2]Graduate school of Material Science, University of Hyogo, 3-2-1 Kouto, Kamigori, Hyogo, 678-1297, Japan

Email*: anh.le@riken.jp

**Abstract**. The aim of this study is to clarify the physics which governs the transition from epsilon phase to zeta phase of solid oxygen observed experimentally at 96 GPa using density functional theory (DFT). The transition was predicted at 40 GPa with PBE functional. Then the Hubbard correction was added to enhance the localization of $p$-orbital of oxygen. The epsilon-zeta transition pressure was significantly improved to 70 GPa. Finally, we included the non-local van der Waals correction. The transition pressure slightly increases to 80 GPa. These results demonstrate that the contribution from Hubbard term is superior to van der Waals term.

## 1. Introduction

Solid oxygen under high pressure has been studied widely for many decades [1-10]. Recently, A. J. Ochoa-Calle *et al* [9] have reported their calculations using various methods from Hartree-Fock (HF) to DFT with LDA, PBE, Meta-GGA, and hybrid functionals in order to obtain a good agreement with the experimental results of the epsilon-zeta transition at 96 GPa [1, 6, 10]. They found that the hybrid functionals yield better quantitative results compared to non-hybrid functionals and HF method. The role of hybrid functionals in this case is apparently still unclear because the agreement was reach when they used 27% of non-local HF exchange for their hybrid functional but HF method with 100% non-local HF exchange predicted no epsilon-zeta phase transition up to 160 GPa.

X-ray diffraction study reported that epsilon-oxygen has a monoclinic *C*2/*m* symmetry in which four $O_2$ molecules associate into a unit cell called $(O_2)_4$ [5]. The electronic properties of epsilon-oxygen thus strongly depends on the localization of electronic states in $(O_2)_4$ and the non-local interaction between the units $(O_2)_4$. Our strategy is as follows: we first conducted the calculation with a conventional DFT functional. Then we included the Hubbard term for improving the localization of oxygen atomic orbitals. Finally, we added the van der Waals correction to improve the description of non-local interaction.

## 2. Computational details

The calculation was performed in the framework of DFT [11] using the Quantum Espresso package [12] which implements the norm-conserving pseudopotential Martins-Troullier method [13]. The number of k-points in the irreducible Brilloin zone was equal to 84 (the 5 x 6 x 8 sampling). The kinetic energy cutoff was set at 150 Ry with $10^{-9}$ Ry total energy convergence. We carried out the variable-cell optimization with Broyden–Fletcher–Goldfarb–Shanno (BFGS) algorithm [14] for both ions and unit cell dynamics. The initial structure of the geometry optimization at 10 GPa was chose based on the X-ray measurement at 17.6 GPa [5]. The convergence threshold on forces for ionic minimization was set at $10^{-3}$ a.u. The $C2/m$ symmetry was assigned to epsilon phase. It was shown that the magnetic properties of solid oxygen collapse in epsilon phase [4]. Thus we only carried non spin-polarized calculation. For the local Hubbard interaction, we employed the rotationally invariant scheme with a simplified version of Cococcioni *et al* [15]:

$$E_U = \frac{U_{eff}}{2} \sum Tr\left[n^{I,\sigma}\left(1-n^{I,\sigma}\right)\right] \quad (1)$$

where $n^{I,\sigma}$ is the on-site occupation matrix of oxygen. Details of $E_U$ and $U_{eff}$ were mentioned in ref. [15]. The Hubbard correction was applied only to the *p*-orbitals. The calculated result of $U_{eff}$ for epsilon phase at 10 GPa was 9.6 eV and virtually did not change at higher pressure up to 50 GPa. The $U_{eff}$ then slightly increased to 10 eV at higher pressure up to 100 GPa. We assumed 9.6 eV is a good value for epsilon and zeta phases at the pressure regime we investigated. We applied 9.6 eV of $U_{eff}$ for Hubbard correction in whole range of pressure from 10 GPa to 140 GPa. To investigate the van der Waals interaction, we used the non-local functional vdW-DF which used revPBE [16] for exchange GGA term [17]. We also tested the vdW-DF2 [18] which used optB86b exchange for GGA term. We found that the vdW-DF functional gives the results closer to the experimental data than vdW-DF2 functional. In this paper, we report the results of vdW-DF functional. Moreover, for comparison, we also used the semi-empirical method DFT-D [19] which is widely used in calculation of van der Waals interaction.

## 3. Results and discussion

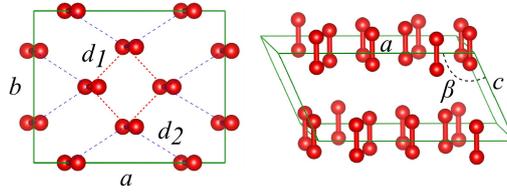

**Figure 1** (color online) the top and bird-eye views of a unit cell of epsilon phase of solid oxygen.

Figure 1 shows the top and bird-eye views of our model for epsilon-oxygen. In this study we investigate the evolution of lattice parameters *a, b, c* with pressure.

### 3.1. The structure calculations with conventional DFT functionals

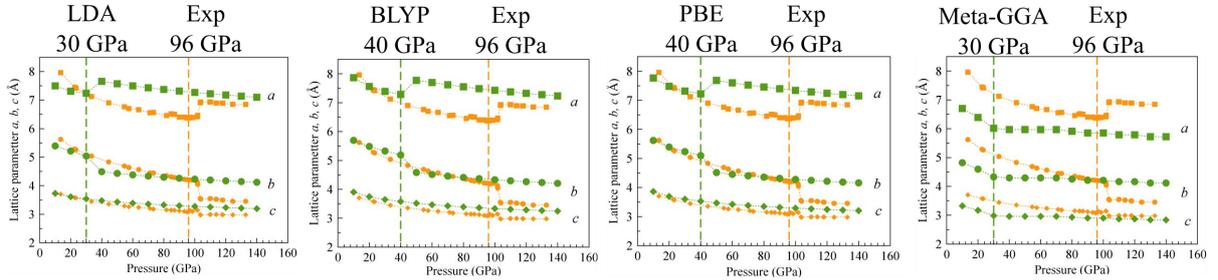

**Figure 2** (color online) The comparison of LDA, BLYP, PBE and Meta-GGA calculations for lattice parameters *a, b, c*. The experimental results are referred from ref. 6 (orange).

In this section, we carried out the variable cell optimization with conventional DFT functionals: LDA, PBE, BLYP, and Meta-GGA. The functionals were chosen based on the Jacob's ladder of exchange-correlation functionals [20]. Figure 2 shows evolution of lattice parameter *a*, *b*, *c* with pressures. For comparison, the experimental data obtained from Weck *et al* [6] are also included. The

LDA yields the epsilon-zeta transition at 30 GPa while the experiment data shows the transition at 96 GPa. The PBE and BLYP functionals improve the LDA's results by 10 GPa, i.e., the transition pressure is at 40 GPa. Interestingly, the Meta-GGA gives the transition at 30 GPa. This demonstrates that another deterministic approach to study epsilon and zeta oxygen should be developed to understand the nature of solid oxygen.

### 3.2. The impact of Hubbard correction

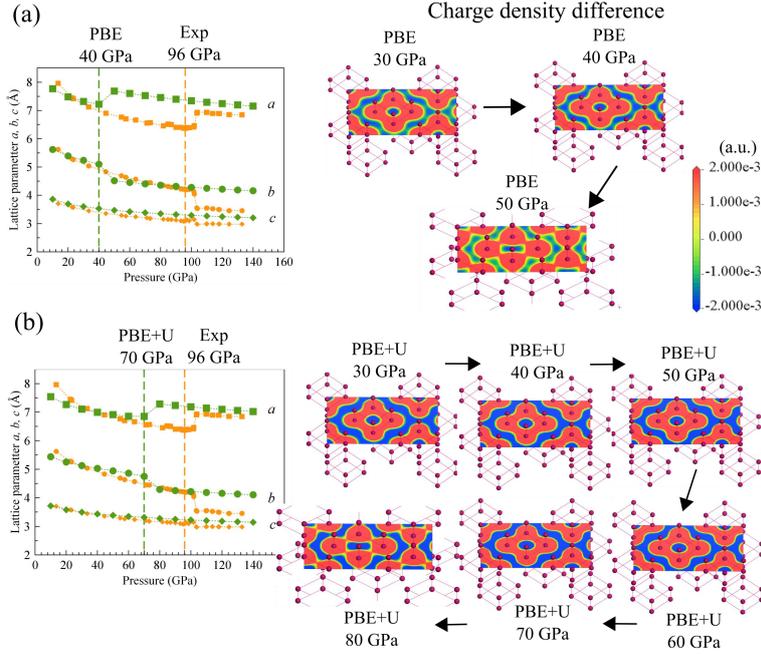

**Figure 3** (color online) Calculated lattice parameters in comparison with experimental measurement in ref. 6: (a) PBE calculation, (b) PBE+U calculation. The right hand side shows the evolution of charge density difference.

To further enhance the accuracy, we study the role of localization of electrons by introducing the Hubbard correction. The Hubbard energy was calculated from Eq. (1). Figure 3(a) and (b) show the lattice parameter $a, b, c$ calculated with PBE and PBE+U, respectively. We can see that the transition pressure is much improved with the presence of U. In particular, the PBE+U shows the transition pressure at 70 GPa. To clarify how U improves the transition pressure, the charge density difference (CDD) was calculated as the charge density minus superposition of atomic densities. The CDD is positive where the bond is formed and is negative where electron is lost. In Fig. 3, we show the cross-section of CDD on the $ab$ plane ($c = 0$). It can be seen from Fig. 3 that the CDD is positive within the region of a cluster $(O_2)_4$ and is negative at the region in between two $(O_2)_4$ clusters. Compared to the PBE calculation, the PBE+U gives the CDD to be more negative at the region between two $(O_2)_4$ clusters. This suggests that U makes electron to be more localized within the $(O_2)_4$ region. Therefore, higher pressure is needed to make these clusters strongly interact each other in order to get the transformation into zeta phase.

### 3.3. The impact of non-local van der Waals interaction

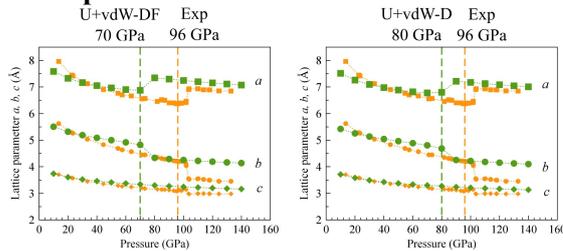

**Figure 4** (color online) The comparison between (a) U+vdW-DF, and (b) U+vdW-D

Recently, there are two approaches to include the van der Waals interaction into DFT calculation: the first approach is modifying a non-local term directly in a functional called vdW-DF functional and calculate the total energy based on this functional, the second one is adding an empirical van der Waals energy to the total energy calculated from a conventional functional such as PBE called vdW-D. In this study, for comparison, we consider both approaches. Figure 4 shows the lattice parameter *a, b, c* calculated from U+vdW-DF, and U+vdW-D method. The U+vdW-DF predicts the transition pressure at 70 GPa while the result for the U+vdW-D is slightly enhanced to 80 GPa. That means the contribution of the van der Waals interaction to the transition is small. This is understandable because, under high pressure, the interatomic distances decrease so that the role of non-local interaction becomes smaller. This suggests that there is a competition between van der Waals and Hubbard term. The competition was also observed between van der Waals interaction and magnetic interactions in solid oxygen [21-23].

**4. Summary and Conclusions**
We carried out the DFT calculations to investigate the transition from epsilon to zeta phase in solid oxygen. The results show that the Hubbard correction strongly improves the prediction of transition pressure to 70 GPa. In contrast, the impact of van der Waals interactions is smaller.

**Acknowledgments** This research was supported by MEXT as "Exploratory Challenge on Post-K computer" (Frontiers of Basic Science: Challenging the Limits) and the RIKEN iTHES Project. This research used the computational resources of the K computer provided by the RIKEN Advanced Institute for Computational Science through the HPCI System Research project (Project ID: hp160251/hp170220) and the Hokusai supercomputer system provided by RIKEN Advanced Center for Computing and Communication.

**References**
[1] S. Desgreniers, Y. K. Vohra, A. L. Ruoff, 1990 J. Phys. Chem., **94**, 1117-1122
[2] Y. Akahama, H. Kawamura *et al*., 1995 Phys. Rev. Lett., **74**, 23 (4690-4693)
[3] G. Weck, P. Loubeyre, R. LeToullec, 2002 Phys. Rev. Lett., **88**, 035504
[4] Yu. A. Freiman, H. J. Jodl, 2004 Physics Reports, **401**, 1-228
[5] L. F. Lundegaard, G. Weck *et al*., 2006 Nature, **443**, 7108 (201-204)
[6] G. Weck, S. Desgreniers *et al*., 2009 Phys. Rev. Lett., **102**, 25
[7] A. J. Ochoa-Calle, C. M. Zicovich-Wilson, and A. Ramírez-Solís, 2015 Phys. Rev. B, **92**, 8
[8] Y. Ma, A. R. Oganov, and C. W. Glass, 2007 Phys. Rev. B, **76**, 064101
[9] A. J. Ochoa-Calle, C. M. Zicovich-Wilson *et al*., 2015 J. Chem. Theory Comput., **11** (3), 1195–1205
[10] H. Fujihisa, Y. Akahama, *et al*., 2006 Phys. Rev. Lett., **97**, 085503
[11] P. Hohenberg and W. Kohn, 1964 Phys. Rev., **136**, B864
[12] P. Giannozzi, S. Baroni, M. Calandra *et al*., 2009 J. Phys.: Condens.Matter, **21**, 395502; The pseudopotentials were downloaded from http://www/quantum-espresso.org/pseudopotentials/
[13] N. Troullier and J. L. Martine, 1991 Phys. Rev. B, **43**, 1993
[14] C. G. Broyden, 1970 J. Inst. Math. Appl., **6**, 76; R. Fletcher, 1970 Comput. J., **13**, 317; D. Goldrarb, 1970 Math. Comp. **24**, 23; D. F. Shanno, 1970 Math. Comp., **24**, 647
[15] M. Cococcioni and S. Gironcoli, 2005 Phys. Rev. B, **71**, 035105
[16] Zhang and Yang, 1998 Phys. Rev. Lett., **80**, 890
[17] M. Dion, H. Rydberg, E. Schröder *et al*., 2004 Phys. Rev. Lett., **92**, 246401
[18] J. Klimeš, D. R. Bowler, and A. Michaelides, 2011 Phys. Rev. B, **83**, 195131
[19] S. Grimme, 2006 J. Comp. Chem., **27** (15), 1787-1799
[20] J. P. Perdew and K. Schmidt, 2001 AIP Conference Proceedings, **577**, 1
[21] S. Kasamatsu, T. Kato, and O. Sugino, 2017 Phys. Rev. B, **95**, 235120
[22] M. Obata, M. Nakamura, *et al*., 2015 J. Phys. Soc. Jpn., **84**, 024715
[23] L. Shi and Y. P. Wang, 2016 Jounal of Magnetism and Magnetic Materials **405**, 1-8